\documentstyle[twoside,fleqn,espcrc2,psfig]{article}

\def\sfrac#1#2{{\textstyle{#1 \over #2}}}
\def\agt{\hbox{${\lower.40ex\hbox{$>$}
\atop \raise.20ex\hbox{$\sim$}}$}}
\def\alt{\hbox{${\lower.40ex\hbox{$<$}
\atop \raise.20ex\hbox{$\sim$}}$}}

\title{Improvement, dynamical fermions, and heavy quark screening
in QCD$_3$}

\author{Howard D. Trottier\address{Physics Department, 
Simon Fraser University, Burnaby, B.C., Canada V5A 1S6, and \\
\ Newman Laboratory of Nuclear Studies, Cornell University, 
Ithaca, NY 14853-5001}}

\begin{document}

\begin{abstract}
First results from simulations of improved actions 
for both gauge fields and staggered fermion fields in 
three dimensional QCD are presented. This work provides insight
into some issues of relevance to lattice theories in 
four dimensions. In particular, the renormalization of 
the bare lattice coupling is dramatically reduced when the 
tree-level $O(a^2)$ improved action is used. 
Naik improvement of the staggered fermion action produces 
little reduction in scaling violations of the rho
meson mass. String breaking in the heavy quark
potential in the unquenched theory is also clearly resolved, using 
Wilson loops to bound the ground state energy.

\end{abstract}


\maketitle

\section*{}

There is a long history of lattice studies of three-dimensional 
QCD, which exhibits much of the physics of the four-dimensional 
theory, including linear confinement \cite{Ambjorn}, and 
spontaneous breakdown of a ``chiral'' symmetry \cite{Kogut}. 

QCD$_3$ is a super-renormalizable theory and, with fermionic 
matter fields, only disconnected vacuum bubble diagrams 
are ultraviolet divergent. In a lattice regularization of QCD$_3$,
nonrenormalizable operators can serve to mimick the effects of 
continuum states that are excluded by the cutoff \cite{GPL}. 
In addition, the bare coupling $g_0^2$ (of dimension $1$) 
and fermion mass $m_0$ must absorb cutoff dependent 
renormalizations. However, these renormalizations 
vanish as $a \to 0$.

The Wilson gluon action in three dimensions has the usual form,
${\cal S}_{\rm Wil} = - \beta \sum_{x,\,\mu>\nu} P_{\mu\nu}$,
where in SU(2) $P_{\mu\nu}$ is one-half the trace of the 
plaquette, and $\beta =  4/(g_0^2a)$.
At $O(a^2)$ the tree-level action differs from the
continuum by the operator
$\sfrac{a^2}{24} \int d^3x \mbox{Tr}F_{\mu\nu}
(D_\mu^2 + D_\nu^2) F_{\mu\nu}$.

Simulations were done here using a tree-level $O(a^2)$-accurate 
anisotropic gluon action \cite{Symanzik}
\begin{equation}
   {\cal S}_{\rm imp} = - \beta \! \sum_{x,\mu>\nu} \!
   \xi_{\mu\nu} \left[
   \sfrac53 P_{\mu\nu}
 - \sfrac{1}{12} (R_{\mu\nu} + R_{\nu\mu})
   \right] ,
\end{equation}
where $R_{\mu\nu}$ is one half the trace of the $2\times1$ 
rectangle in the $\mu\times\nu$~plane, with
$\xi_{3i} = \xi_{i3} = a_s/a_t$ ($a\equiv a_s)$, and
$\xi_{ij} = a_t/a_s$ ($i,j=1,2$). 

The fact that $g_0^2$ and $m_0$ are cutoff-independent in
the continuum limit means that one obtains extremely 
simple scaling laws for physical quantities. In particular,
masses in lattice units (including the input bare
quark masses) should satisfy $\beta (a m) = \mbox{constant}$ 
in the continuum limit.

\begin{figure}[htb]
\vspace{-9pt}
\psfig{file=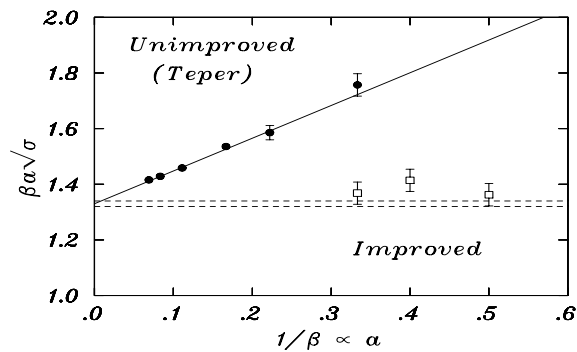,height=4.5cm,width=7.5cm,angle=90}
\vspace{-18pt}
\caption{Lattice spacing dependence of $\beta a \sqrt\sigma$.}
\vspace{-15pt}
\label{fig:sigma}
\end{figure}

\begin{figure}[htb]
\vspace{-9pt}
\psfig{file=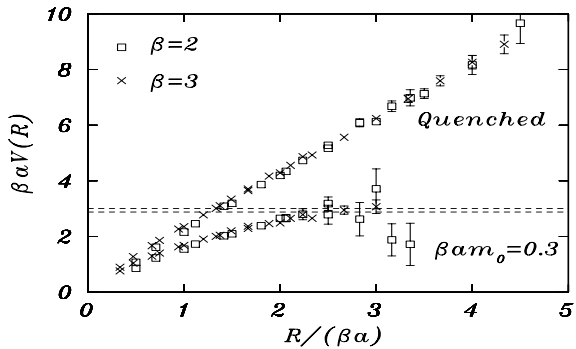,height=4.5cm,width=7.5cm,angle=90}
\vspace{-18pt}
\caption{Static potential for improved glue,
with and without dynamical staggered fermions.}
\vspace{-15pt}
\label{fig:potd}
\end{figure}

Results for the string tension for the unimproved action
were obtained by Teper \cite{Teper}, and are shown
as the solid circles in Fig. \ref{fig:sigma}
(the dashed lines show the continuum limit estimated in
\cite{Teper}, $\beta a \sqrt\sigma = 1.33(1)$).
Results for the improved action obtained here (on lattices 
with $a_t/a_s=1/4$) are shown as the open squares 
in Fig. \ref{fig:sigma}; results for the static potential
are shown in Fig. \ref{fig:potd}. 

A superficially surprising aspect of the results is the 
$O(a)$ scaling violation evident in the Wilson action data 
for $\sqrt\sigma / g_0^2$. In fact this demonstrates the need to
renormalize the bare coupling at finite $a$. 
In four dimensions, it is known that perturbative expansions 
in the bare coupling are spoiled by large corrections; higher 
dimension operators generate $O(a^0)$ renormalizations of 
the coupling, from insertions in ultraviolet-divergent
loop diagrams. A renormalized coupling defined by a physical 
quantity absorbs the scale-independent renormalization, 
and greatly improves perturbative expansions
of many quantities \cite{LepMac}.

In QCD$_3$ higher dimension operators in the lattice 
action can likewise generate a large renormalization of 
$g_0^2$. On the other hand, loop diagrams in QCD$_3$ converge 
as the cutoff is removed.
In particular, one-loop effects are of $O(g_0^2 a)$, 
as required by dimensional analysis. 
This produces the linear scaling violation 
in $\sqrt\sigma / g_0^2$. Teper has done extensive
calculations of 3D glueball masses $m_G$ \cite{Teper}, 
and showed that the ratios $m_G / \sqrt\sigma$ exhibit 
$O(a^2)$ scaling violations, as expected, since 
$\sqrt\sigma$ defines a  renormalized coupling.

The distinctive signature of the renormalization of the bare 
coupling in the 3D theory exposes an interesting feature
of the lattice action. One might expect to see a reduction
in the renormalization of $g_0^2$ when $S_{\rm imp}$ is used,
compared to simulations with the
Wilson action. Remarkably, one finds that the renormalization
is in fact almost completely eliminated; 
the error in $\sqrt\sigma / g_0^2$ at $\beta=2$ 
is reduced from about 45\% with the Wilson action, to less
than about 5\% with $S_{\rm imp}$.

This is a genuinely surprising result, 
since all of the higher dimension 
operators present in $S_{\rm imp}$ should produce a
leading $O(g_0^2 a)$ renormalization of the bare coupling. 
This includes effects of tadpole diagrams induced by higher order 
terms in the link expansion (the tadpole
diagram in 3D is linearly divergent and, for $S_{\rm Wil}$,
$\langle 1 - \sfrac12 \mbox{Tr} U_\mu\rangle_{\rm Landau} 
= 0.063 g_0^2 a + O(g_0^2 a)^2$ );
the results in Figs. \ref{fig:sigma}, \ref{fig:potd} were obtained 
{\it without\/} tadpole renormalization.
Apparently, the operator series in the effective action 
converges very rapidly in 3D, even at scales near 
the lattice cutoff.

Simulations of the static quark potential in the presence of 
dynamical Kogut-Susskind fermions were also done here.
The staggered action in 3D \cite{Burden} is identical in form 
to the 4D action, and describes two flavors of four-component spinors:
\begin{eqnarray}
    {\cal S}_{\rm K-S} = \sum_{x,\mu} \eta_\mu(x)
   \bar\chi(x) \bigl[ U_\mu(x) \chi(x+\hat\mu) \ \ \ \ \ \ & &
\nonumber \\
   - U^\dagger_\mu(x-\hat\mu) \chi(x-\hat\mu) \bigr] 
   + 2 am_0 \sum_x \bar\chi(x) \chi(x) . & &
\end{eqnarray}
Unquenched simulations were done on isotropic lattices at $\beta=2$ 
($12^2\times8$) and at $\beta=3$ ($16^2\times10$).
Hybrid molecular dynamics \cite{Gottlieb} were used, with 
improved glue. The bare quark mass was fixed 
in units of $g_0^2$, with $\beta a m_0 = 0.3$. 
The unquenched heavy-light meson mass \cite{Fiebig} 
was also computed at $\beta=3$, and 
is shown as the dashed lines in Fig. \ref{fig:potd}.

String breaking is clearly demonstrated,
with the unquenched potential approaching twice the 
heavy-light meson mass at large separations. Excellent scaling 
behavior in the potential is also observed (the lattice 
spacing increases by 50\% from $\beta=3$ to $\beta=2$),
despite the fact that $m_0$ is expected to 
undergo a cutoff-dependent renormalization, of 
$O(g_0^2 a)$ at one-loop. Significant scaling violations 
in the unquenched potential were seen when
the Wilson gluon action was used. 

One can roughly estimate the string breaking distance $R_b$ 
in physical units from this data, $R_b / (\beta a) \approx 2.5$. 
If the quenched string tension in the 3D theory 
is equated with its physical value in 4D, 
$\sqrt\sigma \approx 0.44$~GeV,
this implies $R_b \sim 1.5$~fm, which is similar 
to estimates in 4D \cite{Sommer}. 
The spacing at $\beta=2$ can also be estimated 
as $a \sim 0.3$~fm.

Only (fuzzy) Wilson loop operators were used in these
calculations. Little evidence of string breaking has 
previously been seen using Wilson loops,
and it has been suggested that these operators
may have little overlap with the ``broken string'' state
\cite{Gusken}. If true, then working on a coarse
lattice as was done here should prove advantageous, 
as this generally improves the overlap
of an operator with the lowest-lying state. Moreover,
this allows a more efficient probe of the correlation 
function at larger physical times, which is crucial 
in order to isolate the ground state;
at $\beta = 2$ the unquenched effective potential 
$V(R,T)$ reached a plateau only near $T \approx 4 a$,
corresponding to a large ``physical'' time of about 1~fm.

Simulations were also done with some improvement of the 
staggered fermion action. At $O(a^2)$ there
are two sources of discretization error in the 4D 
action: a correction to the kinetics of the staggered field, 
and flavor-changing operators that must be added to cancel 
interactions mediated by highly virtual gluons \cite{Luo}.
Since the 3D action has the same lattice symmetries \cite{Burden},
one expects similar tree-level $O(a^2)$ errors.

Pion and rho meson propagators in quenched  
backgrounds were evaluated for the Kogut-Susskind action, and for
the Naik-improved action, which only corrects for the kinetics
of the fermion field (the Naik action used here is identical 
in form to the 4D theory studied in \cite{MILC}). 
Wall sources and local sinks were used, with two-dimensional 
Coulomb gauge fixing. In order to eliminate possible 
$O(g_0^2 a)$ errors due to a renormalization of
the bare quark mass, chiral extrapolations were done, using 
five quark masses (linear extrapolations for 
$m_\pi^2$ and $m_\rho$ gave good fits).

\begin{figure}[htb]
\vspace{-9pt}
\psfig{file=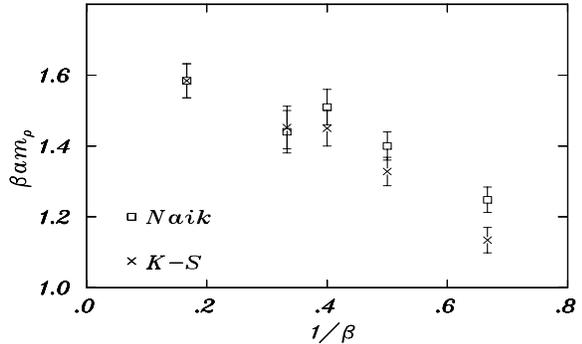,height=4.5cm,width=7.5cm,angle=90}
\vspace{-18pt}
\caption{Quenched rho meson mass.}
\vspace{-15pt}
\label{fig:scaling}
\end{figure}

Results for the $\rho$ mass in the chiral limit are
shown Fig. \ref{fig:scaling}. Simulations 
were done on lattices with $a_t/a_s=1/4$, 
and volumes from $8^2\times16$ at 
$\beta=1.5$ to $24^2 \times 48$ at $\beta=6$.
The Naik action shows little improvement;
tadpole renormalization also makes little difference.
[Given the high degree of improvement of the gluon action, 
one would expect $O(a^2)$ scaling violations in $m_\rho / g_0^2$,
with a sufficiently accurate chiral extrapolation.
Unfortunately, the data is not good enough to discriminate
between $O(a)$ and $O(a^2)$ fits.]

These results suggest that the dominant discretization 
errors in the 3D staggered action come from flavor-changing 
interactions. This is consistent with studies of flavor
symmetry breaking in four dimensions \cite{MILC}.

I am indebted to Peter Lepage for his explanation of
the results reported here, from which much of the
discussion in this paper originates. I also thank 
R. Fiebig, R. Woloshyn and M. Teper for fruitful conversations.

\end{document}